\documentclass[showpacs,preprint,amsmath,amssymb]{revtex4}
\usepackage{graphicx}
\usepackage{dcolumn}
\usepackage{bm}

\begin{document}

\title{Study of transmission properties for waveguide bends by use of a circular photonic crystal}

\author{Sanshui Xiao}
\email{sanshui@imit.kth.se}
\author{Min Qiu}
\affiliation{Laboratory of Optics, Photonics and Quantum
Electronics, Department of Microelectronics and Information
Technology,\\ Royal Institute of Technology (KTH), Electrum 229,
16440 Kista, Sweden}


\begin{abstract}
We study the transmission properties for the waveguide bends
composed by a circular photonic crystal.  Two types (Y and U type)
of the waveguide bends utilizing the circular photonic crystal are
studied. It has been shown, compared with the conventional
photonic crystal waveguide bends, transmission properties for
these bends can be significantly improved. Over a $6.4\%$
bandwidth, less than 1-dB loss/bend are observed. U bent
waveguide, i.e., $180^o$ bend, can be easily realized with low
loss using the circular photonic crystal.
\end{abstract}

\pacs{42.70.Qs, 78.20.Bh, 02.70.Bf} \maketitle

\section{Introduction}
Photonic Crystals (PhCs) are artificial structures with
periodically modulating refractive indices and have photonic
bandgaps wherein the propagation of photons is prohibited
\cite{Yab1987P1,John1987P1,Joannopoulos}. PhCs have received
considerable attention owing to their abilities for the
realization of ultra-compact and multi-functional devices for
high-density photonic integrated circuits (PICs)
\cite{Manolatou1999P1}. It is inevitable for introducing waveguide
bends in the high-density PICs.  This kind of waveguide bends,
specifically sharp corners, represents a discontinuity for a wave
propagating through the waveguide and becomes one of the main
contributions in not only generating loss but also limiting the
bandwidth of the transmitted signal. Various alternative
approaches for bend designs have been theoretically or
experimentally studied, e.g., deforming the PhCs lattice near the
bend \cite{Ntakis2004P1,Talneau2002}, adding a defect at the bend
\cite{Chutinan2002P1}, optimization the waveguide bends
\cite{Jensen2004P1,Borel2004P1,Frandsen2004P1,Miao2004P1,Smajic2003P1}
and using of polycrystalline PC lattice \cite{Sharkawy2003P1}.
Most of these works are based on the optimization method for the
bends, in which the structure of the waveguide bends are always
much complicated.

In this paper, we propose a simple method for designing waveguide
bends and demonstrate the use of a circular photonic crystal as a
waveguide bend to improve transmission property of the bend in a
two-dimensional triangular photonic crystal. Y and U type of
waveguide bends are considered in the following text.

\section{Design and simulation results}
\begin{figure}[htb]
\centering\includegraphics[width=9cm]{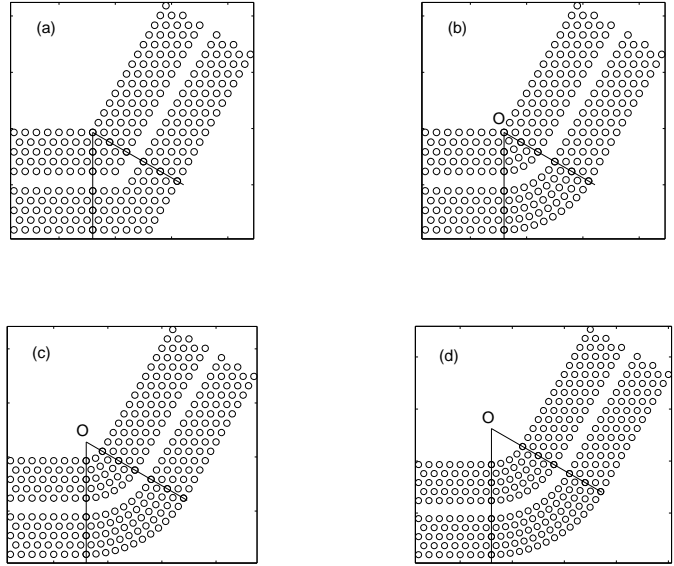}
\caption{\label{Fig1} Schematics of photonic crystal waveguide for
the $60^o$ bend. Generic bend configuration is shown in (a). The
bends using the circular photonic crystal are shown in (b),(c) and
(d). Only difference for three structures is the position of the
center ($O$) for the circular photonic crystal.}
\end{figure}

The two-dimensional photonic crystal we considered here is a
triangular lattice of air holes in a dielectric medium with a
lattice constant $a$ and a hole radius $r=0.3a$. The refractive
index of the dielectric medium 3.24 has been assumed,
corresponding to the the index of InP/GaInAsP around
$\lambda=1.55\mu m$. It can be obtained by the plane-wave
expansion method \cite{Johnson2001P1} that such a structure has a
bandgap for transverse-electric (TE) modes from $0.22 (a/\lambda)$
to $0.29(a/\lambda)$. To guide light, photonic crystal waveguides
(PhCWs) are then introduced by removing one row of air holes,
which are oriented along $\Gamma K$ direction. For the triangular
lattice configuration, a conventional PhCW bend is naturally bent
in steps of $60^o$, which is shown in Fig. \ref{Fig1}(a). Due to
the discontinuity of the generic PhCW bend, light will be
scattered by air holes around the corner when going through it,
which leads to high bending loss and narrow bandwidth of the
transmitted signal. To overcome these limitations, we design the
waveguide bend by use of part of a circular photonic crystal
(CPC)\cite{Horiuchi2004P1} instead of the generic PhC bend, which
is shown in Fig. 1(b). The CPC is non-periodic but systematic
arrangement of air holes, which exhibits sixfold symmetry. The air
holes are arranged in the form of concentric circles with radial
distance $d=\sqrt{3}a/2$, matched with the triangular photonic
crystal. The positions of the air holes in the $xy$ plane for such
a CPC are given by
\begin{eqnarray}
x=dN \sin \left( \frac{2m\pi}{6N} \right), y=dN \cos \left
(\frac{2m\pi}{6N} \right),
\end{eqnarray}
where $N$, $d$, and $m$ denote the number of concentric circles,
the difference of radii of neighboring concentric circles and the
number of air holes ($0\leq m\leq 6N$), respectively. Results of
transmission spectra for the whole CPC ($0\leq N\leq 9$) show that
there exists an isotropic bandgap for TE modes between
$0.22(a/\lambda)$ to $(0.3a/\lambda)$, which makes it possible
that light can be guided by the waveguide of the CPC. For the CPC
is non-periodic, it should be noted that there is no bandgap in
the region of large radial distance, where the CPC is almost like
a square lattice of air  holes in a dielectric medium (without
bandgap for TE modes). Looked again back to Fig. \ref{Fig1} (b),
compared with the conventional PhCW bend (as shown in Fig.
\ref{Fig1} (a)), the CPC makes the bend much smoother,  e. g., the
discontinuity at the bend becomes much smaller. Moreover, we also
keep the symmetry of the corner, which is vital for improvement of
transmission efficiency as described by Jensen \emph{et. al.}
\cite{Jensen2004P1}. In our all simulations, we use a dielectric
waveguide as input and output waveguide and put a detector in the
output waveguide. In order to accurately obtain transmission
spectra of the waveguide bend,  we need to separate the
transmission spectra of the bend from the complicate propagation
loss, as well as the in/out coupling loss when light travels
through the photonic crystal waveguide. Here we use a straight
photonic crystal waveguide as a reference, which has same
propagation length, as well as identical in-coupling and out
coupling mechanisms. The transmission of the bend can then be
defined as the ratio of the output power $P_0$ for the waveguide
bend to the reference power $P_i$ for the corresponding straight
waveguide, which is given by $T=10log_{10}(P_i/P_0)$. Numerical
simulations for the bends are performed by the two-dimensional
finite-difference-time-domain (FDTD) computational method
\cite{TafloveFDTD} with a boundary treatment of perfectly matched
layers \cite{Berenger1994P1}.

First we study the case of Y type waveguide bend, i.e., $60^o$
bend, in a W1 channel waveguide oriented along $\Gamma K$
direction. The conventional bend in the triangular photonic
crystal is shown in Fig. \ref{Fig1} (a). It has been shown by
several authors
\cite{Ntakis2004P1,Frandsen2004P1,Miao2004P1,Smajic2003P1} that
the transmission of this generic waveguide bend is quite small.
Light will be strongly scattered by air holes around the corner
due to the mismatch of the guided mode. In order to minimize the
scattering, we use a matched circular photonic crystal to connect
the conventional PhCW, which are shown in Fig. \ref{Fig1} (b),(c)
and (d). The regions between two solid lines are the transition
region for the bends.  Only difference for these three structures
is the position of the center ($O$) for the CPC. Waveguides in the
CPC can be denoted by $N$, where $N=5$, $7$ and $9$ represent the
waveguide structures of Fig. \ref{Fig1} (b), (c) and (d),
respectively. Transmission spectra of the bend, i.e, bend loss,
can be obtained by the method mentioned before in order to
eliminate the coupling and the propagation loss.
\begin{figure}[htb]
\centering\includegraphics[width=9cm]{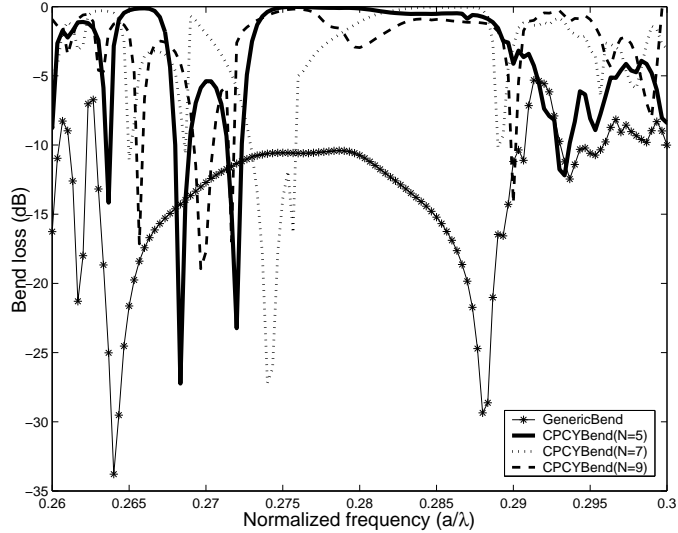}
\caption{\label{Fig2} Bend loss for the waveguide bends. The solid
line with asterisk markers represents the loss for the generic
PhCW bend. The solid, dotted and dashed line represent the loss
for the CPC waveguide structure shown in Fig. 1(b), (c) and (d),
respectively. All spectra have been normalized to the transmission
through straight PhCWs of the same length to eliminate the
coupling and the propagation loss in straight waveguide.}
\end{figure}
Bend losses for the structures shown in Fig. \ref{Fig1} are
plotted in Fig. \ref{Fig2}. The solid line with asterisk markers
represents the bend loss for the conventional PhCW bend and the
solid, dotted and dashed line represent the bend loss for the
three waveguide bends utilizing the CPC, respectively. It is
obviously seen that the bend loss for the generic bend is much
large in our considered frequency region, which is in agreement
with other results
\cite{Ntakis2004P1,Frandsen2004P1,Miao2004P1,Smajic2003P1}. For
the bends by use of the CPC, the bend losses are quite small in a
large frequency domain, which agrees well with what we expected
above.  It can also be seen from Fig. \ref{Fig2} that less than
1-dB loss/bend can be obtained in a normalized frequency range
from $0.272 (a/\lambda)$ to $0.288 (a/\lambda)$. As for the
working wavelength $\lambda=1.55\mu m$, the bandwidth is about $90
nm$. Compared other corresponding work in the literatures
\cite{Borel2004P1,Frandsen2004P1,Miao2004P1}, although the
bandwidth for the low bend loss in this paper is not the best
result, the bandwidth is still enough wide for the practical
application. Moreover, the structure of the waveguide bend is much
simpler than those in Ref.
\cite{Borel2004P1,Frandsen2004P1,Miao2004P1}.

\begin{figure}[htb]
\centering\includegraphics[width=15cm]{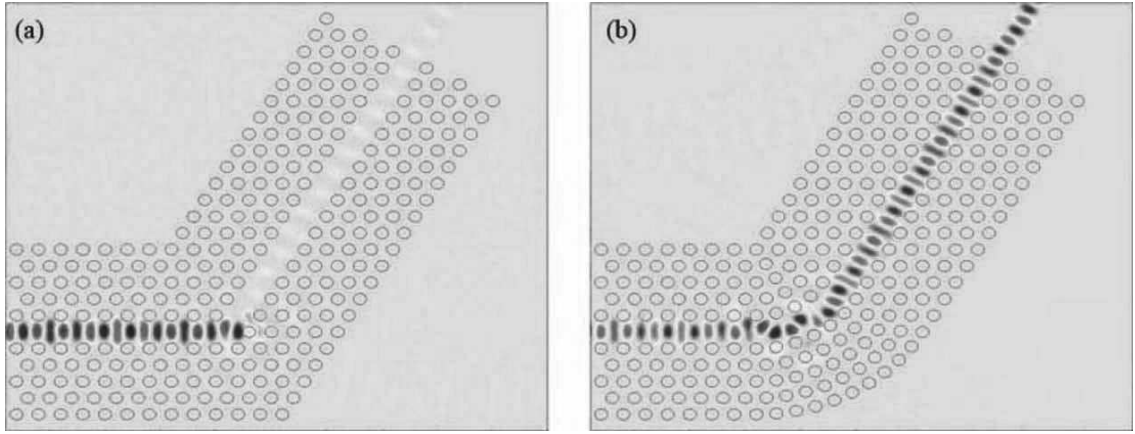}
\caption{\label{Fig3} Steady-state magnetic field distribution
($\omega=0.276 (a/\lambda$)) for the waveguide bends for (a) mode
profile through the generic PhCW bend; (b) mode profile through
the waveguide bend utilizing the CPC, the corresponding structure
is shown in Fig. 1(b).}
\end{figure}
Figure \ref{Fig3} shows the steady-state magnetic field
distribution ($\omega=0.276 (a/\lambda$)) of the mode profile when
light goes through the waveguide bends. The left image shows the
mode behavior for light travelling through the generic PhCW bend
and right image is for the bend by use of the CPC, the
corresponding structure of which is shown in Fig. \ref{Fig1}(b).
One can clearly see from the Fig. \ref{Fig3}(a) that transmission
is quite small for such a generic PhCW bend. However, shown in
Fig. \ref{Fig3}(b), the CPC waveguide bend guides the light
nicely, which is in agreement with the description above.

\begin{figure}[htb]
\centering\includegraphics[width=9cm]{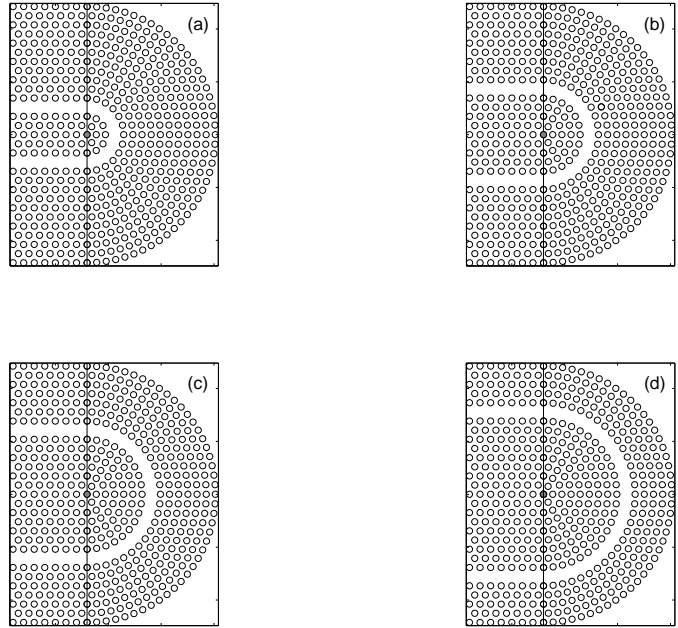}
\caption{\label{Fig4} Schematics of U type photonic crystal
waveguide bends by use of the circular photonic crystal. Left
regions of solid lines are the half circular photonic crystal and
the right are the conventional triangular photonic crystal. The
filled circles are the center circles of the CPC. }
\end{figure}

\begin{figure}[htb]
\centering\includegraphics[width=9cm]{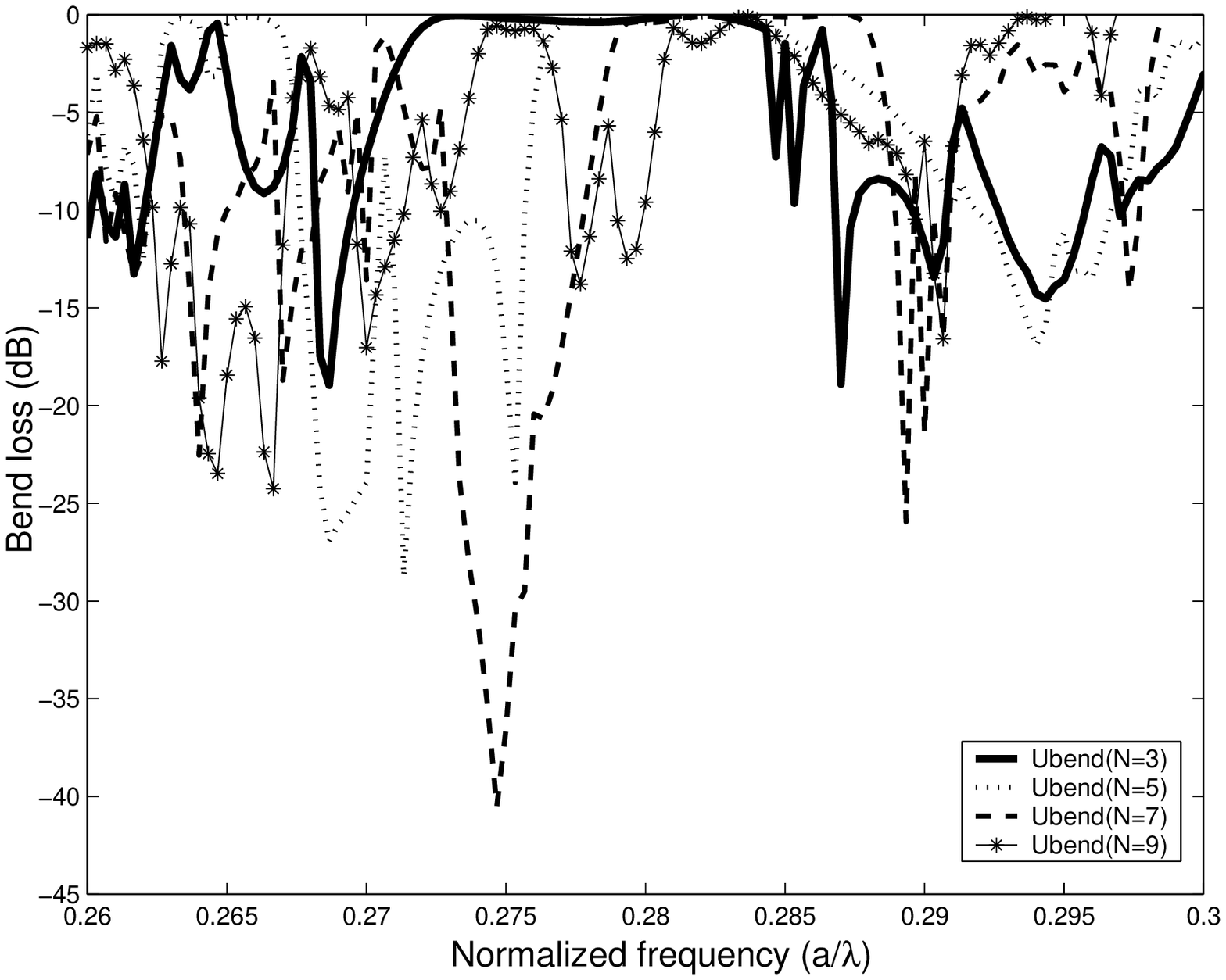}
\caption{\label{Fig5} Bend losses for four different U type
waveguide bends by use of the circular photonic crystal, whose
corresponding structures are shown in Fig. 4. The solid, dotted
and dashed line represent the loss of CPC waveguide structures
shown in Fig. 4(a), (b) and (c), respectively. The solid with
asterisk markers represents the loss for the structure of Fig. 4
(d). }
\end{figure}
As for the high-density, it is necessary to introduce different
type of waveguide bend in PICs.  Next we consider U type photonic
crystal waveguide bend, i.e., $180^o$ bend, by use of the circular
photonic crystal. The CPC bend structure can be introduced by
removing one row of air holes with same radial length, which are
shown in Fig. \ref{Fig4}. Left regions for solid lines in Fig.
\ref{Fig4} are the half circular photonic crystal and the right
are the conventional triangular photonic crystal. The filled
circles are the center circles of the CPC.  The CPC waveguide can
also be denoted by $N=3$, $5$, $7$ and $9$ for four kinds of U
type waveguide bends. Our simulation results are shown in Fig.
\ref{Fig5}. The solid, dotted, dashed line and solid line with
asterisk markers represent the bend loss for the four bends shown
in Fig. \ref{Fig4}, respectively. One can see from Fig.
\ref{Fig5}, the bend loss are quite small in a relative wide band
for these U type waveguide bends. Over $6.4\%$ bandwidth, less
than 1-dB loss/bend can be observed for the structure shown in
Fig. \ref{Fig4} (a). As for the working wavelength
$\lambda=1.55\mu m$, the bandwidth is about $100 nm$, which is
enough wide for application. It can be also seen from Fig.
\ref{Fig5} that the bandwidth for large transmission becomes
narrow, especially for the results of the dotted line, with
increasing of the bend radius. This is mainly caused by the effect
that the bandgap for the CPC will shrink with the increase of the
radial length due to its non-periodic structure.
\begin{figure}[htb]
\centering\includegraphics[width=15cm]{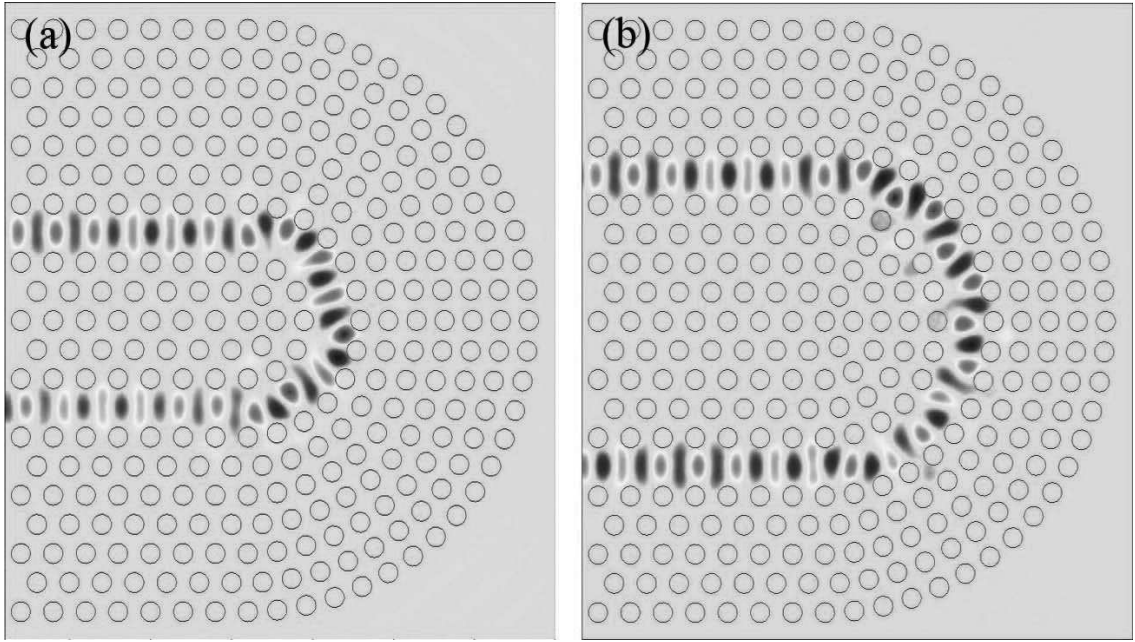}
\caption{\label{Fig6} Steady-state magnetic field distribution
($\omega=0.276 (a/\lambda$)) for the waveguide bends. The
corresponding structure of the bends are shown in Fig. 5 (a) and
Fig. 5 (b). }
\end{figure}
Figure \ref{Fig6} shows the steady-state magnetic field
distribution for the light $\omega=0.2827 (a/\lambda$). The mode
behavior for light travelling through the bend (corresponding
structure is shown in Fig. \ref{Fig5} (a)) is shown in the left
image. Right image shows the mode behavior for light going through
the structure shown in Fig. \ref{Fig5} (b). They both show that
such these CPC waveguide bends can realize $180^o$ turning with
much low bend losses. For the structure of Fig. \ref{Fig5}(a), the
bend radius is about $1 \mu m$ for the working wavelength
$\lambda=1.55\mu m$, which is quite small.

\section{Conclusion}
In conclusion, we have studied the waveguide bends by use of a
circular photonic crystal. Compared with the conventional generic
PhCW bends, the bends with CPC have shown good transmission
properties not only of the transmission efficiency, but also of
the bandwidth with the large transmission. Two types (Y type and U
type) of the waveguide bends utilizing the circular photonic
crystal are considered in this paper. Over a $6.4\%$ bandwidth,
less than 1-dB loss/bend are observed. For the working wavelength
$\lambda=1.55\mu m$, the bandwidth is about $100 nm$, which is
enough wide for the application. Moreover, very small bend radius
of about $1\mu m$ for U type bend can be easily obtained. Further
work on experimental verification of our results mentioned above
has been in the progress, which will be presented in the future.

\section{Acknowledgments}
This work was supported by the Swedish Foundation for Strategic
Research (SSF) on INGVAR program, the SSF Strategic Research
Center in Photonics, and the Swedish Research Council (VR) under
project 2003-5501.


\begin{thebibliography}{16}
\expandafter\ifx\csname
natexlab\endcsname\relax\def\natexlab#1{#1}\fi
\expandafter\ifx\csname bibnamefont\endcsname\relax
  \def\bibnamefont#1{#1}\fi
\expandafter\ifx\csname bibfnamefont\endcsname\relax
  \def\bibfnamefont#1{#1}\fi
\expandafter\ifx\csname citenamefont\endcsname\relax
  \def\citenamefont#1{#1}\fi
\expandafter\ifx\csname url\endcsname\relax
  \def\url#1{\texttt{#1}}\fi
\expandafter\ifx\csname
urlprefix\endcsname\relax\def\urlprefix{URL }\fi
\providecommand{\bibinfo}[2]{#2}
\providecommand{\eprint}[2][]{\url{#2}}

\bibitem[{\citenamefont{Yablonovitch}(1987)}]{Yab1987P1}
\bibinfo{author}{\bibfnamefont{E.}~\bibnamefont{Yablonovitch}},
  \bibinfo{journal}{Phys. Rev. Lett.} \textbf{\bibinfo{volume}{58}},
  \bibinfo{pages}{2059} (\bibinfo{year}{1987}).

\bibitem[{\citenamefont{John}(1987)}]{John1987P1}
\bibinfo{author}{\bibfnamefont{S.}~\bibnamefont{John}}, \bibinfo{journal}{Phys.
  Rev. Lett.} \textbf{\bibinfo{volume}{58}}, \bibinfo{pages}{2486}
  (\bibinfo{year}{1987}).

\bibitem[{\citenamefont{Joannopoulos et~al.}(1995)\citenamefont{Joannopoulos,
  Meade, and Winn}}]{Joannopoulos}
\bibinfo{author}{\bibfnamefont{J.~D.} \bibnamefont{Joannopoulos}},
  \bibinfo{author}{\bibfnamefont{R.~D.} \bibnamefont{Meade}}, \bibnamefont{and}
  \bibinfo{author}{\bibfnamefont{J.}~\bibnamefont{Winn}},
  \emph{\bibinfo{title}{Photonic Crystals: Modling the Flow of Light}}
  (\bibinfo{publisher}{Princeton Univ. Press}, \bibinfo{address}{Princeton,
  NJ}, \bibinfo{year}{1995}), \bibinfo{edition}{1st} ed.

\bibitem[{\citenamefont{Manolatou et~al.}(1999)\citenamefont{Manolatou,
  Johnson, Fan, Villenueve, Haus, and Joannopoulos}}]{Manolatou1999P1}
\bibinfo{author}{\bibfnamefont{C.}~\bibnamefont{Manolatou}},
  \bibinfo{author}{\bibfnamefont{S.~G.} \bibnamefont{Johnson}},
  \bibinfo{author}{\bibfnamefont{S.}~\bibnamefont{Fan}},
  \bibinfo{author}{\bibfnamefont{P.~R.} \bibnamefont{Villenueve}},
  \bibinfo{author}{\bibfnamefont{H.~A.} \bibnamefont{Haus}}, \bibnamefont{and}
  \bibinfo{author}{\bibfnamefont{J.~D.} \bibnamefont{Joannopoulos}},
  \bibinfo{journal}{J. Lightwave Technol.} \textbf{\bibinfo{volume}{17}},
  \bibinfo{pages}{1682} (\bibinfo{year}{1999}).

\bibitem[{\citenamefont{Ntakis et~al.}(2004)\citenamefont{Ntakis, Pottier, and
  De~La~Rue}}]{Ntakis2004P1}
\bibinfo{author}{\bibfnamefont{I.}~\bibnamefont{Ntakis}},
  \bibinfo{author}{\bibfnamefont{P.}~\bibnamefont{Pottier}}, \bibnamefont{and}
  \bibinfo{author}{\bibfnamefont{R.~M.} \bibnamefont{De~La~Rue}},
  \bibinfo{journal}{J. Appl. Phys.} \textbf{\bibinfo{volume}{96}},
  \bibinfo{pages}{12} (\bibinfo{year}{2004}).

\bibitem{Talneau2002}
\bibinfo{author}{\bibfnamefont{A.}~\bibnamefont{Talneau}},
  \bibinfo{author}{\bibfnamefont{L.}~\bibnamefont{Le~Gouezigou}},
  \bibinfo{author}{\bibfnamefont{N.}~\bibnamefont{Bouadma}},
  \bibinfo{author}{\bibfnamefont{M.}~\bibnamefont{Kafesaki}},
  \bibinfo{author}{\bibfnamefont{C. ~M.}~\bibnamefont{Soukoulis}},
  \bibnamefont{and}
  \bibinfo{author}{\bibfnamefont{M.} \bibnamefont{Agio}},
  \bibinfo{journal}{Appl. Phys. Lett.} \textbf{\bibinfo{volume}{80}},
  \bibinfo{pages}{547} (\bibinfo{year}{2002}).


\bibitem[{\citenamefont{Chutinan et~al.}(2002)\citenamefont{Chutinan, Okano,
  and Noda}}]{Chutinan2002P1}
\bibinfo{author}{\bibfnamefont{A.}~\bibnamefont{Chutinan}},
  \bibinfo{author}{\bibfnamefont{M.}~\bibnamefont{Okano}}, \bibnamefont{and}
  \bibinfo{author}{\bibfnamefont{S.}~\bibnamefont{Noda}},
  \bibinfo{journal}{Appl. Phys. Lett.} \textbf{\bibinfo{volume}{80}},
  \bibinfo{pages}{1698} (\bibinfo{year}{2002}).

\bibitem[{\citenamefont{Jensen and Sigmund}(2004)}]{Jensen2004P1}
\bibinfo{author}{\bibfnamefont{J.~S.} \bibnamefont{Jensen}} \bibnamefont{and}
  \bibinfo{author}{\bibfnamefont{O.}~\bibnamefont{Sigmund}},
  \bibinfo{journal}{Appl. Phys. Lett.} \textbf{\bibinfo{volume}{84}},
  \bibinfo{pages}{2022} (\bibinfo{year}{2004}).

\bibitem[{\citenamefont{Borel et~al.}(2004)\citenamefont{Borel, Harpoth,
  Frandsen, and Kristensen}}]{Borel2004P1}
\bibinfo{author}{\bibfnamefont{P.~I.} \bibnamefont{Borel}},
  \bibinfo{author}{\bibfnamefont{A.}~\bibnamefont{Harpoth}},
  \bibinfo{author}{\bibfnamefont{L.~H.} \bibnamefont{Frandsen}},
  \bibnamefont{and}
  \bibinfo{author}{\bibfnamefont{M.}~\bibnamefont{Kristensen}},
  \bibinfo{journal}{Opt. Express} \textbf{\bibinfo{volume}{12}},
  \bibinfo{pages}{1996} (\bibinfo{year}{2004}).

\bibitem[{\citenamefont{Frandsen et~al.}(2004)\citenamefont{Frandsen, Harpoth,
  Borel, Kristensen, Jensen, and Sigmund}}]{Frandsen2004P1}
\bibinfo{author}{\bibfnamefont{L.~H.} \bibnamefont{Frandsen}},
  \bibinfo{author}{\bibfnamefont{A.}~\bibnamefont{Harpoth}},
  \bibinfo{author}{\bibfnamefont{P.~I.} \bibnamefont{Borel}},
  \bibinfo{author}{\bibfnamefont{M.}~\bibnamefont{Kristensen}},
  \bibinfo{author}{\bibfnamefont{J.~S.} \bibnamefont{Jensen}},
  \bibnamefont{and} \bibinfo{author}{\bibfnamefont{O.}~\bibnamefont{Sigmund}},
  \bibinfo{journal}{Opt. Express} \textbf{\bibinfo{volume}{12}},
  \bibinfo{pages}{5916} (\bibinfo{year}{2004}).

\bibitem[{\citenamefont{Miao et~al.}(2004)\citenamefont{Miao, Chen, Shi,
  Murakowski, and Prather}}]{Miao2004P1}
\bibinfo{author}{\bibfnamefont{B.}~\bibnamefont{Miao}},
  \bibinfo{author}{\bibfnamefont{C.}~\bibnamefont{Chen}},
  \bibinfo{author}{\bibfnamefont{S.}~\bibnamefont{Shi}},
  \bibinfo{author}{\bibfnamefont{J.}~\bibnamefont{Murakowski}},
  \bibnamefont{and} \bibinfo{author}{\bibfnamefont{D.~W.}
  \bibnamefont{Prather}}, \bibinfo{journal}{IEEE Photon. Tech. Lett.}
  \textbf{\bibinfo{volume}{16}}, \bibinfo{pages}{2469} (\bibinfo{year}{2004}).

\bibitem[{\citenamefont{Smajic et~al.}(2003)\citenamefont{Smajic, Hafner, and
  Erni}}]{Smajic2003P1}
\bibinfo{author}{\bibfnamefont{J.}~\bibnamefont{Smajic}},
  \bibinfo{author}{\bibfnamefont{C.}~\bibnamefont{Hafner}}, \bibnamefont{and}
  \bibinfo{author}{\bibfnamefont{D.}~\bibnamefont{Erni}},
  \bibinfo{journal}{Phys. Rev. Lett.} \textbf{\bibinfo{volume}{11}},
  \bibinfo{pages}{1378} (\bibinfo{year}{2003}).

\bibitem[{\citenamefont{Sharkawy et~al.}(2003)\citenamefont{Sharkawy, Pustai,
  Shi, and Prather}}]{Sharkawy2003P1}
\bibinfo{author}{\bibfnamefont{A.}~\bibnamefont{Sharkawy}},
  \bibinfo{author}{\bibfnamefont{D.}~\bibnamefont{Pustai}},
  \bibinfo{author}{\bibfnamefont{S.}~\bibnamefont{Shi}}, \bibnamefont{and}
  \bibinfo{author}{\bibfnamefont{D.~W.} \bibnamefont{Prather}},
  \bibinfo{journal}{Opt. Lett.} \textbf{\bibinfo{volume}{28}},
  \bibinfo{pages}{1197} (\bibinfo{year}{2003}).

\bibitem[{\citenamefont{Johnson and Joannopoulos}(2001)}]{Johnson2001P1}
\bibinfo{author}{\bibfnamefont{S.~G.} \bibnamefont{Johnson}} \bibnamefont{and}
  \bibinfo{author}{\bibfnamefont{J.~D.} \bibnamefont{Joannopoulos}},
  \bibinfo{journal}{Opt. Express} \textbf{\bibinfo{volume}{8}},
  \bibinfo{pages}{173} (\bibinfo{year}{2001}).

\bibitem[{\citenamefont{Horiuchi et~al.}(2004)\citenamefont{Horiuchi, Segawa,
  Nozokido, Mizumo, and Miyazaki}}]{Horiuchi2004P1}
\bibinfo{author}{\bibfnamefont{N.}~\bibnamefont{Horiuchi}},
  \bibinfo{author}{\bibfnamefont{Y.}~\bibnamefont{Segawa}},
  \bibinfo{author}{\bibfnamefont{T.}~\bibnamefont{Nozokido}},
  \bibinfo{author}{\bibfnamefont{K.}~\bibnamefont{Mizumo}}, \bibnamefont{and}
  \bibinfo{author}{\bibfnamefont{H.}~\bibnamefont{Miyazaki}},
  \bibinfo{journal}{Opt. Lett.} \textbf{\bibinfo{volume}{29}},
  \bibinfo{pages}{1084} (\bibinfo{year}{2004}).

\bibitem[{\citenamefont{Taflove}(2000)}]{TafloveFDTD}
\bibinfo{author}{\bibfnamefont{A.}~\bibnamefont{Taflove}},
  \emph{\bibinfo{title}{Computational Electrodynamics: The Finite-Difference
  Time-Domain Method}} (\bibinfo{publisher}{Artech House INC},
  \bibinfo{address}{Norwood}, \bibinfo{year}{2000}), \bibinfo{edition}{2nd} ed.

\bibitem[{\citenamefont{Berenger}(1994)}]{Berenger1994P1}
\bibinfo{author}{\bibfnamefont{J.~P.} \bibnamefont{Berenger}},
  \bibinfo{journal}{J. Comput. Phys.} \textbf{\bibinfo{volume}{114}},
  \bibinfo{pages}{185} (\bibinfo{year}{1994}).

\end{thebibliography}
\end{document}